\documentclass[11pt]{article} 
\usepackage{mystyle-new}
\usepackage{authblk}
\usepackage[normalem]{ulem}
\usepackage[T1]{fontenc}
\usepackage{epsfig,amsmath} 
\usepackage{hepnames,hepunits}
\usepackage{hyperref}
\usepackage{color}
\usepackage{graphicx}
\usepackage{orcidlink}

\definecolor{red}{rgb}{1,0,0}
\def\lesssim{\ \hbox{\raise 2pt \hbox{$<$} \kern -13pt
                     \lower 3pt \hbox{$\sim$}}\ }
\def\greatersim{\ \hbox{\raise 2pt \hbox{$>$} \kern -13pt
                     \lower 3pt \hbox{$\sim$}}\ }

\def\lsim{\mathrel{\rlap{\lower4pt\hbox{\hskip1pt$\sim$}}
    \raise1pt\hbox{$<$}}}                
\def\gsim{\mathrel{\rlap{\lower4pt\hbox{\hskip1pt$\sim$}}
    \raise1pt\hbox{$>$}}}                

\input epsf.tex
\def\desepsf(#1 width #2){\epsfxsize=#2 \epsfbox{#1}}
\def\kt{\ensuremath{k_{\rm T}}}

\def\qt{\ensuremath{q_{\rm t}}}
\def\zdyn{\ensuremath{z_{\rm dyn}}}
\def\zM{\ensuremath{z_{ M}}}

\newcommand{\Pmax}{\mu^2}

\newcommand{\PBM}{PB}

\newcommand{\TMDplotter}{{\sc TMDplotter}}

  \newenvironment{defl}[1]%
  {\begin{list}{}{\settowidth{\labelwidth}{#1}%
  \setlength{\leftmargin}{\labelwidth}%
  \addtolength{\leftmargin}{\labelsep}%
  \setlength{\itemsep}{0pt plus 1pt}
  \setlength{\parsep}{0pt plus 1pt}
  \setlength{\topsep}{0pt plus 1pt}
  \setlength{\partopsep}{0pt plus 1pt}
  \setlength{\parskip}{2mm plus 1mm minus 1mm}
  }}%
  {\end{list}}

\usepackage{amsmath,bm}
\usepackage{lineno}

\usepackage{cite,./mcite}
\usepackage{tikz}
\usepackage[symbol]{footmisc}

\newcommand{\ccfm}{Ciafaloni:1987ur,Catani:1989yc,Catani:1989sg,Marchesini:1994wr}

\newcommand{\dglap}{Gribov:1972ri,Lipatov:1974qm,Altarelli:1977zs,Dokshitzer:1977sg}
\newcommand{\smallxref}{Marchesini:1992jw,Marchesini:1990zy}
\newcommand{\pbref}{Hautmann:2017fcj,Hautmann:2017xtx}

\newcommand{\nnloSplit}{vanNeerven:2000wp,Moch:2004pa,Vogt:2004mw,Vermaseren:2005qc,Blumlein:2021enk,Blumlein:2022gpp,ABLINGER2014263,Ablinger:2017tan,Moch:2014sna,Behring:2019tus,Blumlein:2021ryt}

\def\SMALLX{{\sc Smallx}}
\def\qcdnum{{\sc QCDnum}}

\providecommand{\DOI}[1]{\href{http://dx.doi.org/#1}}

\def\updfevolv{{\sc uPDFevolv2}} 
\newcommand{\updfevolvversion}{2.5.03}

\newcommand{\as}{\ensuremath{\alpha_s}}

\begin{document}

\title{
The Parton Branching evolution package \updfevolv  \footnote{This article is dedicated the memory to M. Botje, author of the \qcdnum\ package.}}
\author[1,2,3]{H.~Jung \thanks{hannes.jung@desy.de}\orcidlink{0000-0002-2964-9845}}
\affil[1]{Deutsches Elektronen-Synchrotron DESY, Germany}
\affil[2]{II. Institut f\"ur Theoretische Physik, Universit\"at Hamburg, Germany}
\affil[3]{Elementary Particle Physics, University of Antwerp, Belgium}
\author[3]{A.~Lelek\thanks{aleksandra.lelek@uantwerpen.be}\orcidlink{0000-0001-5862-2775}}
\author[1]{K.~Moral~Figueroa\thanks{keila.moral.figueroa@desy.de}\orcidlink{0000-0003-1987-1554}}
\author[1]{S.~Taheri~Monfared\thanks{sara.taheri.monfared@desy.de}\orcidlink{0000-0003-2988-7859}}

\date{}
\begin{titlepage} 
\maketitle
\vspace*{-10cm}
\begin{flushright}
DESY-24-083\\
\today
\vspace*{9cm}
\end{flushright}

\begin{abstract}

\updfevolv\ is a software package designed for evolving collinear and Transverse Momentum Dependent (TMD) parton densities using the DGLAP evolution equation. 
A comprehensive description of both the theoretical framework and technical implementation is given, accompanied by a detailed guide on program usage, focusing on customizable parameters.

This report is a technical release note for \updfevolv\ version  \updfevolvversion .
\end{abstract}

\end{titlepage}

{\sffamily\large\bfseries PROGRAM SUMMARY} \\ \\
{\em Title of Program:} \updfevolv\ \updfevolvversion\ \\
{\em Computer for which the program is designed and others on which it is
operable:}   Any with standard Fortran 77 (gfortran) and C++, tested on
                 Linux, MAC.\\
{\em Programming Language used:}  FORTRAN 77, C++ \\
{\em High-speed storage required:}  No \\
{\em Separate documentation available: } No 
{\em Nature of physical problem:}
The evolution equations for parton densities cannot be solved analytically and numerical methods need to be applied. 
Transverse Momentum Dependent parton (TMD) densities can be obtained from the inclusive DGLAP evolution equations once the evolution scale $q'$ is associated with the transverse momentum \qt\ of the emitted parton. \\
{\em Method of solution:} 
The evolution equations for parton densities are solved numerically using a formulation involving Sudakov form factors. The integral equations are of Fredholm type and can be solved iteratively using a Monte Carlo technique. 
The iterative solution allows for a treatment of the kinematic relations at each individual branching process, and thus allows directly to calculate the transverse momentum of the emitted partons, leading to a direct calculation of the TMD parton densities.
\\ 
{\em Restrictions on the complexity of the problem:}   None
\\
{\em Other Program used:}  \qcdnum\ for splitting functions and \as . {\sc Root} for plotting the result.  \\
{\em Download of the program:} \verb+https://updfevolv.hepforge.org+\\
{\em Unusual features of the program:}   None \\
\newpage

\section{Introduction}

\updfevolv\ is a versatile software tool rooted in the \SMALLX~\cite{\smallxref} program\footnote{The \SMALLX\ source code is available from Ref.~\cite{SmallxProgram}}  and built upon the Parton Branching (\PBM ) method~\cite{\pbref}. It serves as a robust platform for evolving parton densities, both collinear and Transverse Momentum Dependent (TMD), utilizing the widely-used DGLAP evolution equation~\cite{\dglap}. With its foundation in \SMALLX\ and the innovative PB method, \updfevolv\ offers researchers a powerful framework for studying the intricate dynamics of parton evolution in high-energy physics.

The \updfevolv\  package described here is a significant extension of {\sc uPDFevolv1}~\cite{Hautmann:2014uua}, which is based on the evolution of gluon densities using the CCFM evolution equation~\cite{\ccfm}.

\section{Theoretical Input}

\subsection{The Parton Branching solution of the DGLAP equation \label{PB_DGLAP-softgluon}}

The DGLAP evolution equation can be solved with the PB method, as detailed in Refs.~\cite{Hautmann:2017fcj, Hautmann:2017xtx}.

We begin by expressing the DGLAP evolution equation for the momentum-weighted parton density \(xf_a(x,\mu^2)\) of parton \(a\) with momentum fraction \(x\) at scale \(\mu\):
\begin{equation}
\label{EvolEq}
 \mu^2 \frac{\partial (x f_a(x,\mu^2))}{\partial \mu^2} = \sum_b \int_x^{1} dz \; P_{ab} \left(z,\as\right) \; \frac{x}{z} f_b\left(\frac{x}{z}, \mu^2\right),
\end{equation}
where \(P_{ab}\) represents the regularized DGLAP splitting functions governing the transition of parton \(b\) into parton \(a\). The function \(P_{ab}\) can be decomposed as follows (following Ref.~\cite{Hautmann:2017fcj}):
\begin{equation}
P_{ab}(z, \alpha_s) = D_{ab}(\alpha_s)\delta(1-z) + K_{ab}(\alpha_s)\frac{1}{(1-z)_{+}} + R_{ab}(z, \alpha_s).
\label{Eq:Pdecomp}
\end{equation}
Here, \(D_{ab}\) and \(K_{ab}\) are coefficients expressed as \(D_{ab}(\alpha_s) = \delta_{ab}d_a(\alpha_s)\) and \(K_{ab}(\alpha_s) = \delta_{ab}k_a(\alpha_s)\), respectively, while \(R_{ab}\) encompasses terms non-singular as \(z \rightarrow 1\). Each coefficient can be expanded in powers of \(\alpha_s\):
\begin{equation}
d_a(\alpha_s) = \sum_{n=1}^{\infty} \left(\frac{\alpha_s}{2\pi}\right)^n d_a^{(n-1)}, \quad
k_a(\alpha_s) = \sum_{n=1}^{\infty} \left(\frac{\alpha_s}{2\pi}\right)^n k_a^{(n-1)}, \quad
R_{ab}(z, \alpha_s) = \sum_{n=1}^{\infty} \left(\frac{\alpha_s}{2\pi}\right)^n R_{ab}^{(n-1)}(z).
\end{equation}
Introducing the real-emission branching probabilities \(P_{ab}^{(R)}(z,\as)\):
\begin{equation}
P_{ab}^{(R)}(z,\as) = K_{ab}(\as) \frac{1}{1-z} + R_{ab}(z,\as) \;,
\label{realPab}
\end{equation}
the solution to the evolution equation for the momentum-weighted parton density \(xf_a(x,\mu^2)\) at scale \(\mu\) is given by:
\begin{equation}
\label{sudintegral2}
x f_a(x,\mu^2) = \Delta_a(\mu^2) x f_a(x,\mu^2_0) + \sum_b \int_{\mu^2_0}^{\mu^2} \frac{dq^{\prime 2}}{q^{\prime 2}} \frac{\Delta_a(\mu^2)}{\Delta_a(q^{\prime 2})} \int_x^{\zM} dz \;  P^{(R)}_{ab}(z, \as) \frac{x}{z} f_b\left(\frac{x}{z}, q^{\prime 2}\right),
\end{equation}
where  \(\mu_0\) is the starting scale, $\Delta_a(\mu^2) := \Delta_a(\mu^2, \mu_0^2)$ is the Sudakov form factor and  ${\bf q}^{\prime}$ is a 2-dimensional vector with ${\bf q}^{\prime \,2} = q^{\prime 2}$.

\subsubsection{Determination of inclusive Parton Densities using the Parton-Branching Method}

The approach outlined above has been employed to determine collinear (\(k_T\)-integrated) parton densities using Monte Carlo techniques with details provided in ~\cite{Hautmann:2017fcj} and briefly summarised in Section ~\ref{MCsolution}.

\begin{figure}[htb]
\centering
\includegraphics[width=\textwidth]{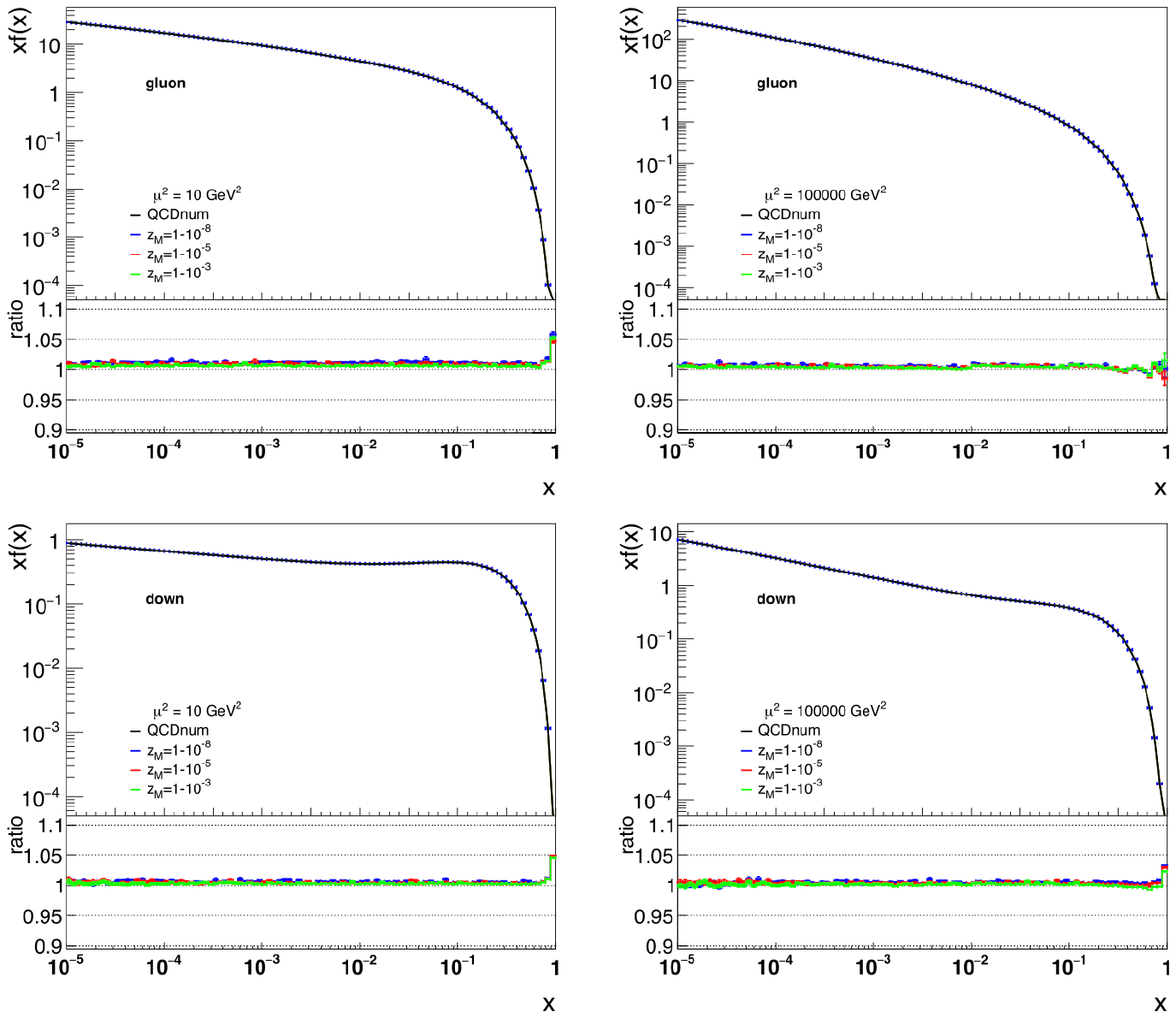}
\caption{Integrated gluon and down-quark distributions at \(\mu^2 = 10\) GeV\(^2\) (left column) and \(\mu^2 = 10^5\) GeV\(^2\) (right column) obtained from the PB solution for different values of \(z_M\), compared with the result from \qcdnum. The ratio plots show the ratio of the results obtained with the PB method to the result from \qcdnum. Figure taken from \protect\cite{Hautmann:2017fcj}.}
\label{Fig:PBzlimit}
\end{figure}

Figure~\ref{Fig:PBzlimit} shows predictions of parton densities evolved to a large scale using the \PBM\ method and compares them with calculations obtained using the conventional tool, \qcdnum ~\cite{Botje:2010ay}. The \PBM\ predictions are presented for various values of \(z_M\). Notably, when \(z_M\) is sufficiently large, the predictions align precisely with semi-analytical calculations, as seen in prior studies \cite{GolecBiernat:2007pu,GolecBiernat:2006xw,Jadach:2007qa,Jadach:2009gm,Jadach:2003bu}.

These findings hold significance in two respects:
\begin{itemize}
\item the DGLAP evolution equation, solved with the concept of resolvable branchings, faithfully reproduces other DGLAP solutions when the "soft resolution" scale \(z_M\) is sufficiently large.
\item An iterative solution of the DGLAP equation using Monte Carlo techniques based on resolvable branchings, such as the PB method, is equivalent to other DGLAP solutions (e.g. \qcdnum).
\end{itemize}

The  PB method has been applied to determine collinear parton densities by fitting them to inclusive DIS measurements \cite{Martinez:2018jxt}. This resulted in two sets of densities, PB-NLO-HERAI+II-set1 and PB-NLO-HERAI+II-Set2, depending on which scale in \(\alpha_s\) is used. In Set1, the evolution scale $ q'$ is utilized, while in Set2, the transverse momentum $q' (1-z)$ is employed.

\begin{figure}[h!tb]
\begin{center} 
\includegraphics[width=0.32\textwidth]{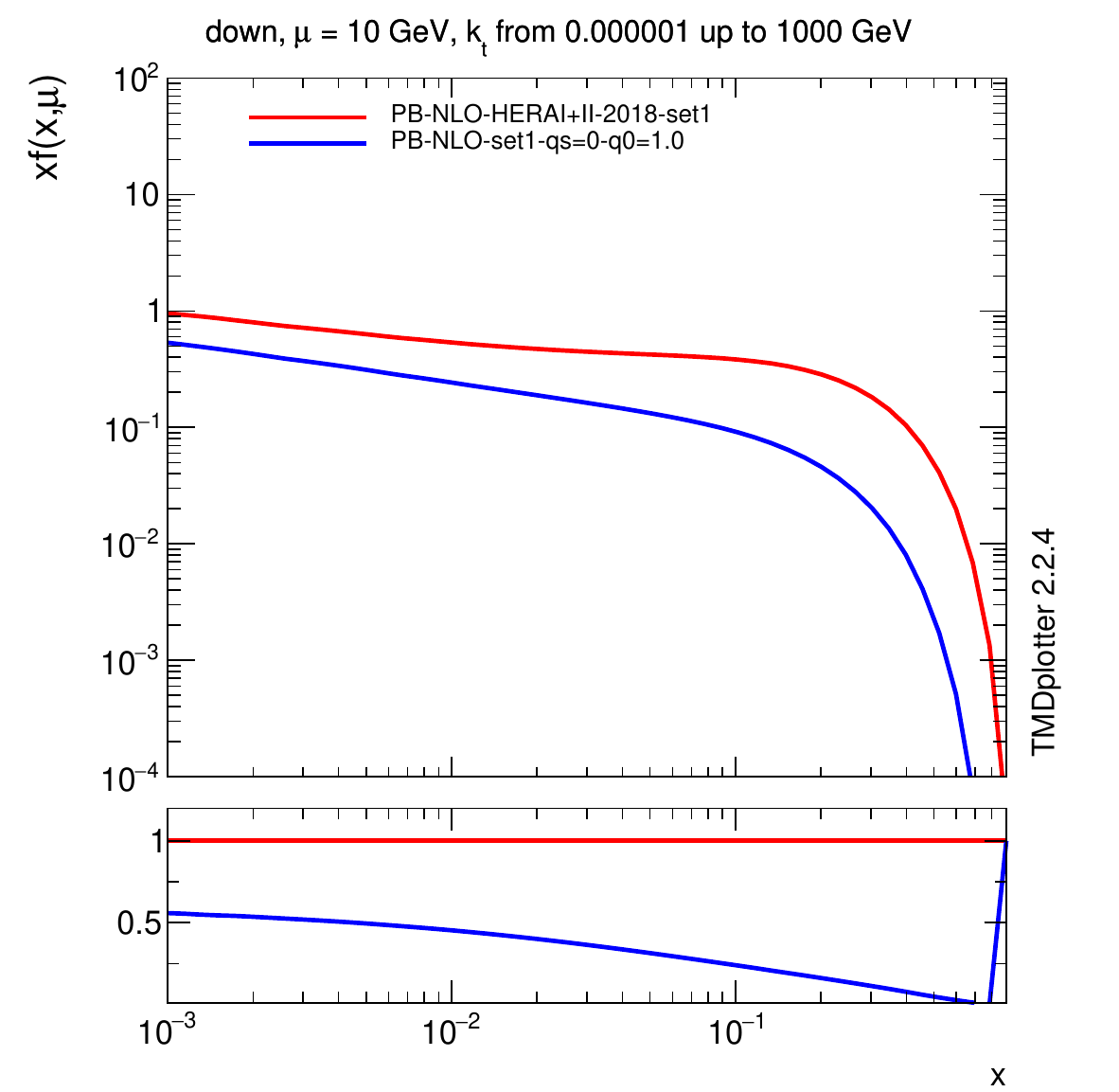}
\includegraphics[width=0.32\textwidth]{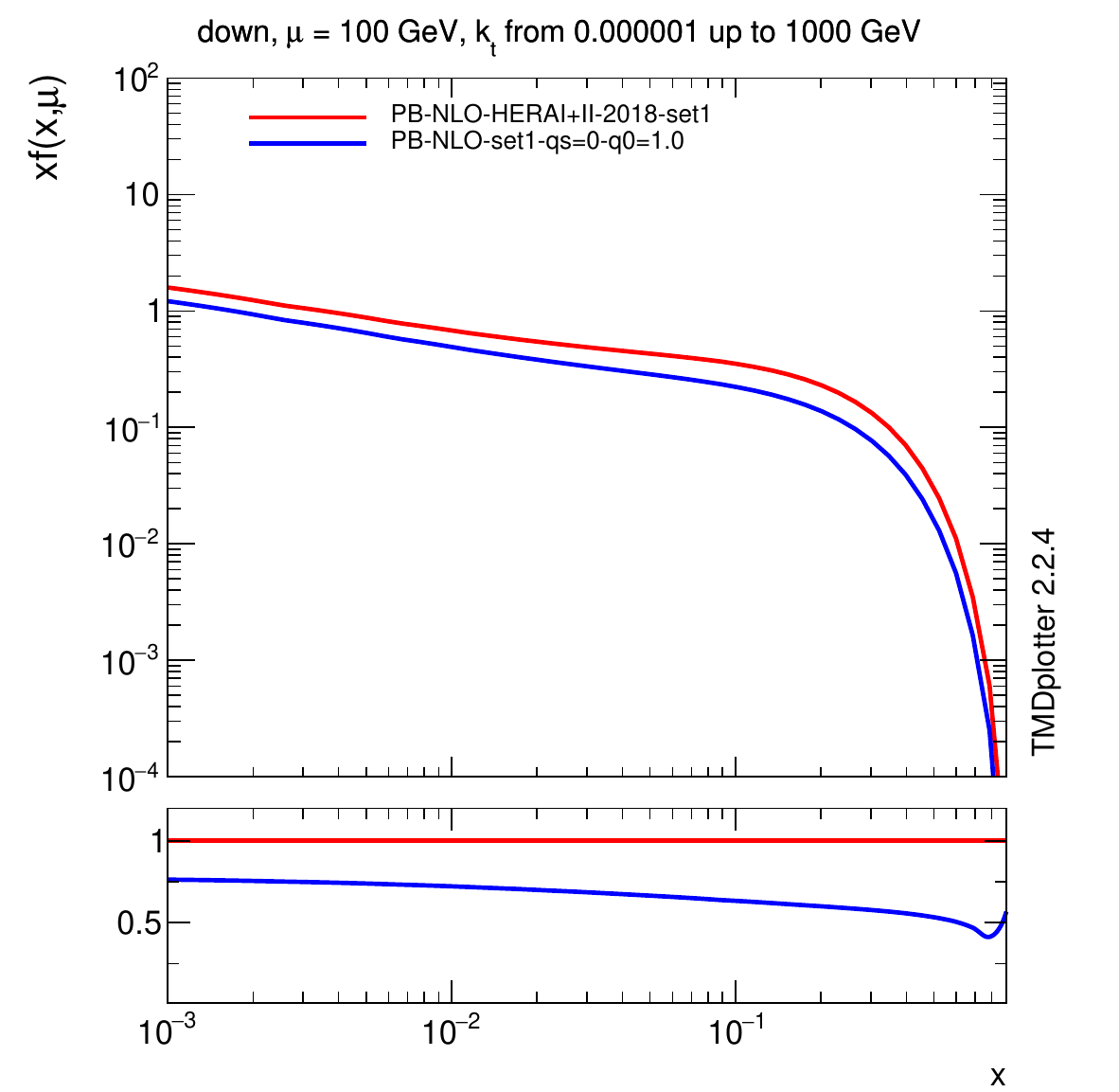}
\includegraphics[width=0.32\textwidth]{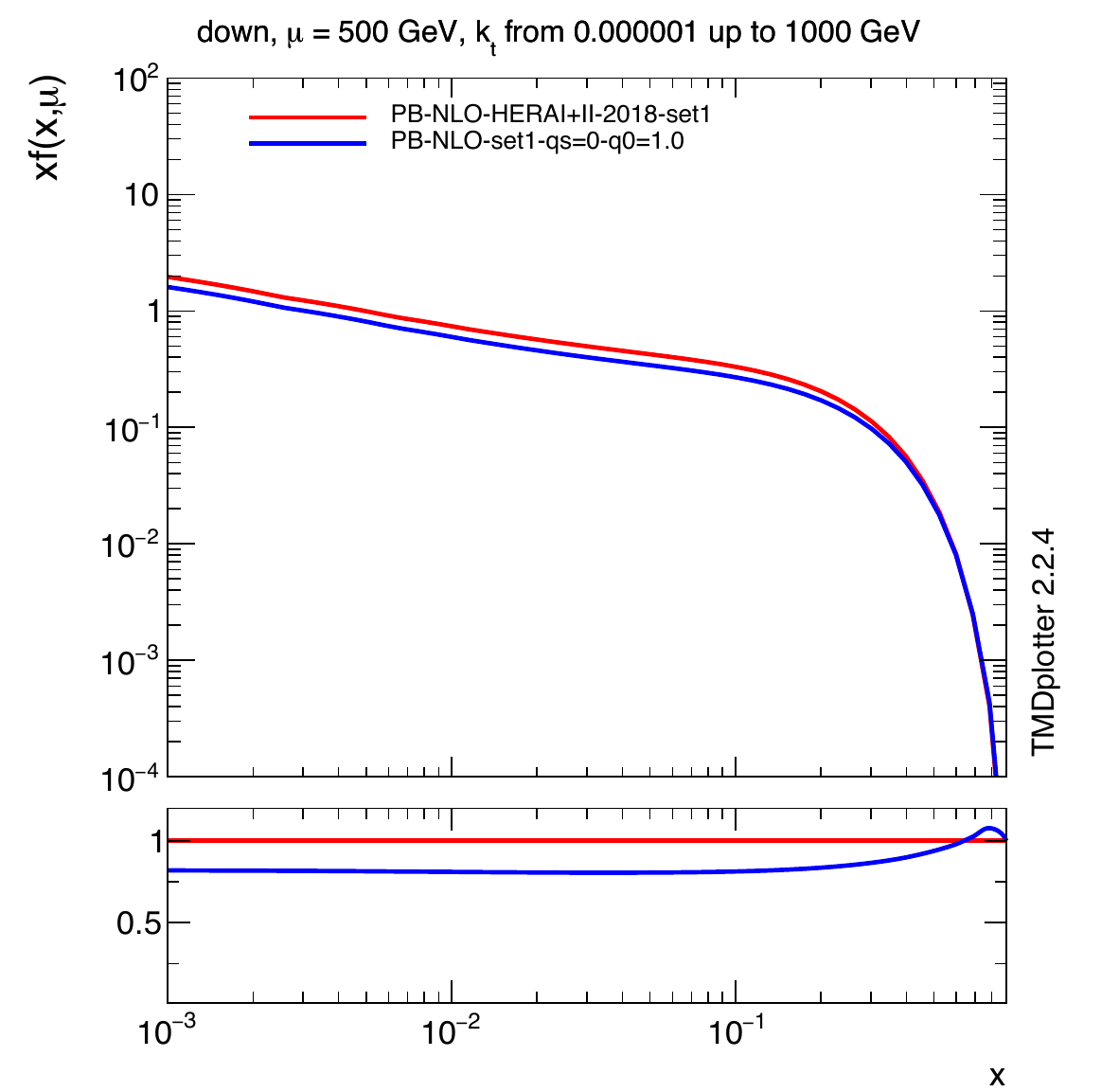}
\caption{\small Integrated down-quark distributions at \(\mu = 10\), \(100\), and \(500\) GeV obtained from the \PBM\ approach for different values of \(z_M\): PB-NLO-HERAI+II-set1 applies \(z_M \to 1\), and PB-NLO-set1-qs=0-q0=1.0 applies \(z_M=\zdyn\) with \(q_0=1\) GeV. The ratio plots show the ratios to the one for \(z_M \to 1\) (figures from \protect\cite{Mendizabal:2023mel}).}
\label{PB-DGLAP}
\end{center}
\end{figure}

It is crucial to note the significance of \(z_M\): it must be close to one to avoid omitting a significant part of the parton evolution.  Here, we illustrate the effect of $z_M = \zdyn = 1 - q_0/q'$, with \(q_0\) being a free resolution parameter of the order of a GeV, on inclusive collinear distributions. The results, obtained using the same parameters and starting distributions as in \cite{Martinez:2018jxt} but with a different cut on \(z_M\), are shown in Figure ~\ref{PB-DGLAP} (for details, refer to \cite{Mendizabal:2023mel}).

Figure~\ref{PB-DGLAP} depicts distributions for down-quarks at different scales, demonstrating the impact when \(z_M\) is dynamical: soft gluons, with transverse momenta even below a resolution scale of \(q_0 = 1 \) GeV, play a significant role in collinear distributions and cannot be neglected.

\subsection{Transverse Momentum Dependent parton densities: Parton Branching TMD}
Below, we present an extension of the DGLAP evolution equation incorporating the dependence on transverse momenta, as outlined by the PB method \cite{Hautmann:2017fcj}. Solving the evolution equation iteratively offers the advantage of treating each splitting explicitly, allowing the application of kinematic relations in every branching, akin to a parton shower process. Consequently, parton distributions can be obtained not only depending on \(x\) and \(\mu\) (as in \(f(x, \mu^2)\)), but also on the transverse momentum ${\bf k}$  of the propagating parton (as in Transverse Momentum Dependent (TMD) parton distributions \({\cal A}(x, {\bf k}, \mu)\)).

\subsubsection{The PB-TMD evolution equation}

We can now extend the DGLAP evolution equation to incorporate the dependence on transverse momenta as described by the PB method \cite{Hautmann:2017fcj}. The extended evolution equation for the transverse momentum dependent parton density ${\cal A}(x,{\bf k},\mu^2) $ is given by:
\begin{eqnarray}
\label{dglap-sudakov-TMD}
  {x {\cal A}}_a(x,{\bf k}, \mu^2) 
 &=&  
\Delta_a (  \mu^2  ) \ 
 {x {\cal A}}_a(x,{\bf k},\mu^2_0)  
 + \sum_b 
\int
{{d^2 {\bf q}^{\prime } } 
\over {\pi {\bf q}^{\prime 2} } }
 \ 
{
{\Delta_a (  \mu^2  )} 
 \over 
{\Delta_a (  {\bf q}^{\prime 2}  
 ) }
}
\ \Theta(\mu^2-{\bf q}^{\prime 2}) \  
\Theta({\bf q}^{\prime 2} - \mu^2_0)
 \nonumber\\ 
&\times&  
\int_x^{\zM} {dz} \;  
P_{ab}^{(R)} (z, \as) 
\;\frac{x}{z} {\cal A}_b\left(\frac{x}{z}, {\bf k}+(1-z) {\bf q}^\prime , 
{\bf q}^{\prime 2}\right)  
  \;\;  ,     
\end{eqnarray}
where the transverse momentum vectors (2-dimensional) ${\bf k}$ and ${\bf q}$ are used to fully account for the transverse momentum dependence. Here we implicitly assume angular ordering (the default option in \updfevolv ) which relates the transverse momentum of the emitted parton $\qt$ to the evolution scale $q'$ via $\qt^2 = (1 - z)^2 q^{\prime\, 2} $.
The final transverse momentum of the propagating parton is calculated as the vectorial sum over intrinsic transverse momentum of the initial parton and all transverse momenta of the emitted partons $i$:
\begin{equation}
{\bf k} = {\bf k}_0- \sum_i {\bf q}_{t,i} \ . 
\end{equation}
 This enables the determination of the corresponding Transverse Momentum Dependent (TMD) parton distribution ${\cal A}(x, {\bf k}, \mu^2)$, in addition to the inclusive distribution $f(x, \mu^2)$, integrated over ${\bf k}$:
\begin{equation}
\int {\cal A}(x,{\bf k},\mu^2) d^2{\bf k}   = f(x,\mu^2) \ .
\end{equation}

Similarly to equation (\ref{sudintegral2}), also equation (\ref{dglap-sudakov-TMD}) can be solved iteratively using a Monte Carlo method ~\cite{Hautmann:2017fcj}.

\begin{figure}[h!tb]
\begin{center} 
\includegraphics[width=0.32\textwidth]{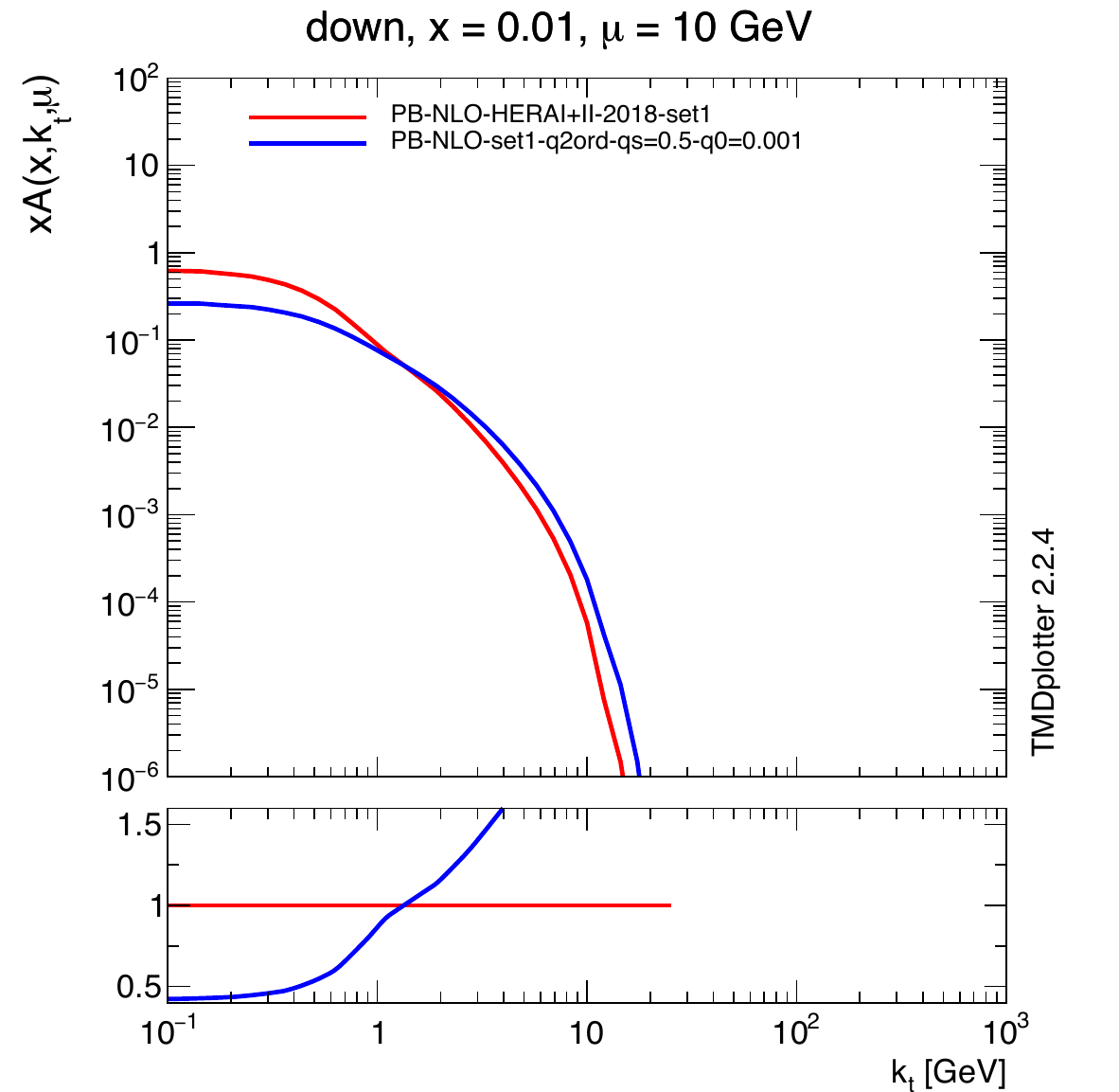}
\includegraphics[width=0.32\textwidth]{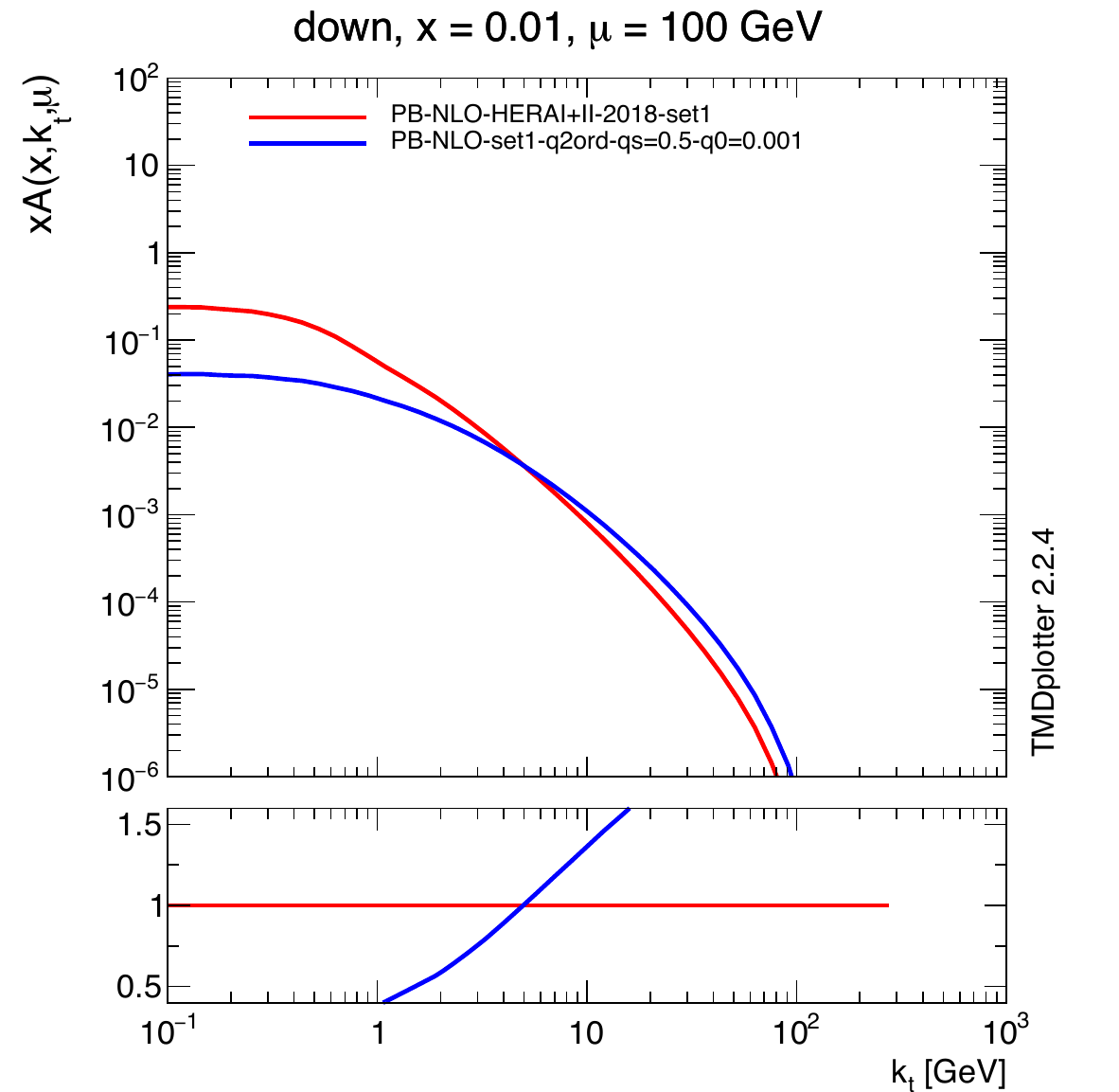}
\includegraphics[width=0.32\textwidth]{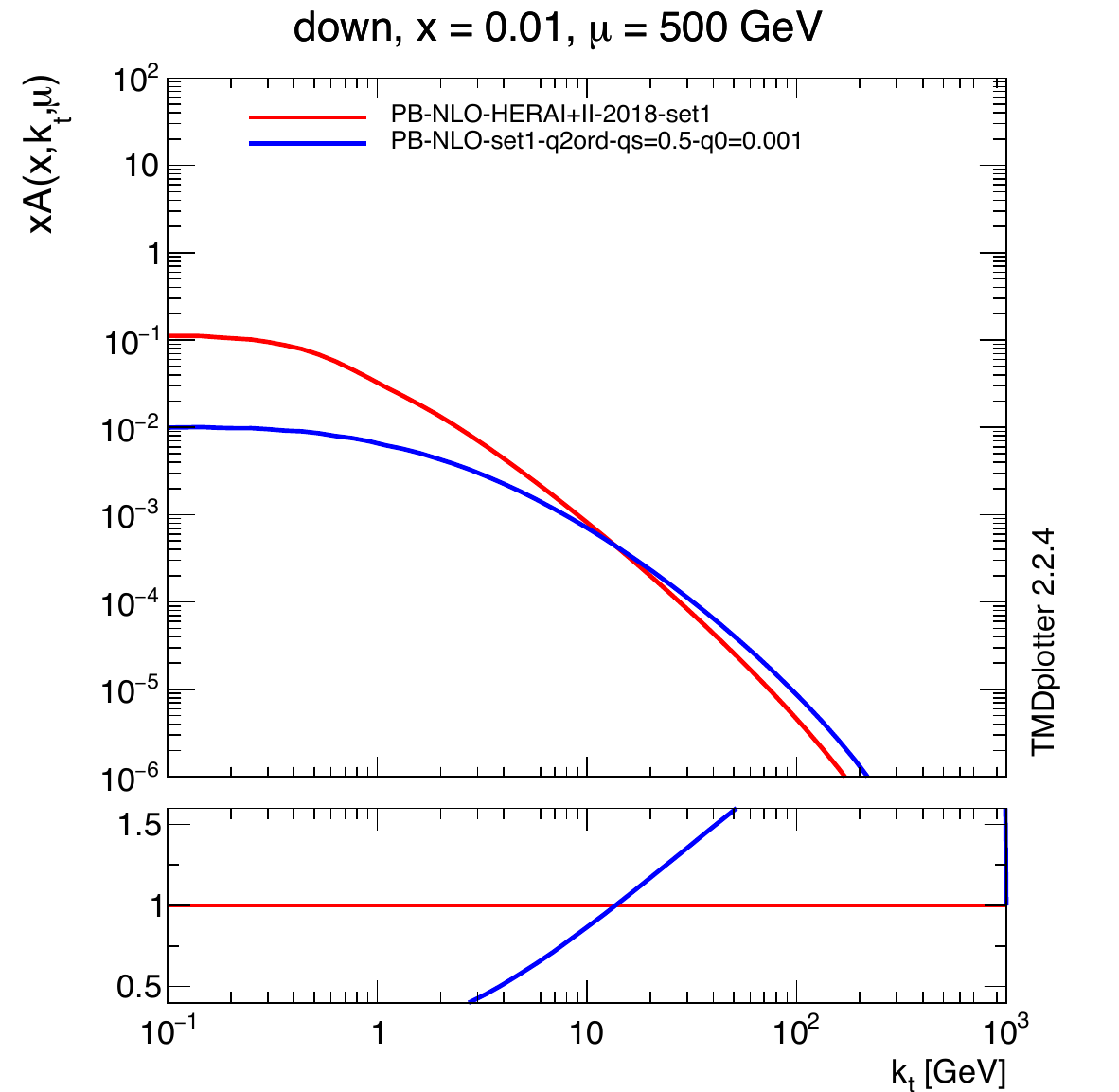}
\caption{\small Transverse momentum distributions of down quarks at $\mu  = 10, 100$~GeV (left, middle column) and $\mu = 500$ GeV (right column). The red curve shows  PB-NLO-HERAI+II-set1, with angular ordering condition applied to relate branching scale $q^{\prime 2}$  and the emitted transverse momentum $q_{t}^2$, the blue curve shows a prediction obtained with virtuality ordering.
 }
\label{TMD-DGLAP}
\end{center}
\end{figure}

In the literature (e.g.~\cite{Bierlich:2022pfr}{p.69}), also virtuality ordering is being discussed, which gives a slightly different relation between the evolution scale and the transverse momentum $\qt^2 = (1 - z) q^{\prime\, 2} $ (available in \updfevolv\ as option \verb+Iqord=1+).  In Figure~\ref{TMD-DGLAP} the transverse momentum distribution for d-quarks is shown for angular ordering (default) as well as for virtuality ordering for different values of the evolution scale $\mu$. It is interesting to observe differences, while the integrated distributions are identical.

\subsubsection{Monte Carlo solution of the  evolution equations}
\label{MCsolution}
As described above, the evolution equations (\ref{sudintegral2}, \ref{dglap-sudakov-TMD}) are  integral equations of the Fredholm type
$$
f(x) = f_0(x) + \lambda \int_a^b K(x,y) f(y)
dy
$$ and can be solved by iteration  as a Neumann series 
\begin{equation}
f(t) = \lim_{n\to \infty} \sum_{i=0}^n \lambda^i u_i(t) \;\; , 
\end{equation}
where 
\begin{eqnarray}
u_0(t) & = &  f_0(t)   \;\;  , \nonumber \\
u_1(t) & = &  \int_a^b K(t,y) f_0(y) dy \;\;  , \nonumber \\
u_2(t) & = &  \int_a^b \int_a^b K(t,y_1)  K(y_1,y_2)f_0(y_2)  dy_2 dy_1      \;\;  ,   \nonumber \\
\cdots  \nonumber  \\
\vdots \nonumber  \\
u_n(t) & = &  \int_a^b \int_a^b \int_a^b  K(t,y_1)  \cdots K(y_{n-1},y_n) f_0(y_n) dy_n \cdots dy_2 dy_1   \;\;  . 
\end{eqnarray}

\begin{figure}[htbp]
\centering \includegraphics[width=0.95\textwidth]{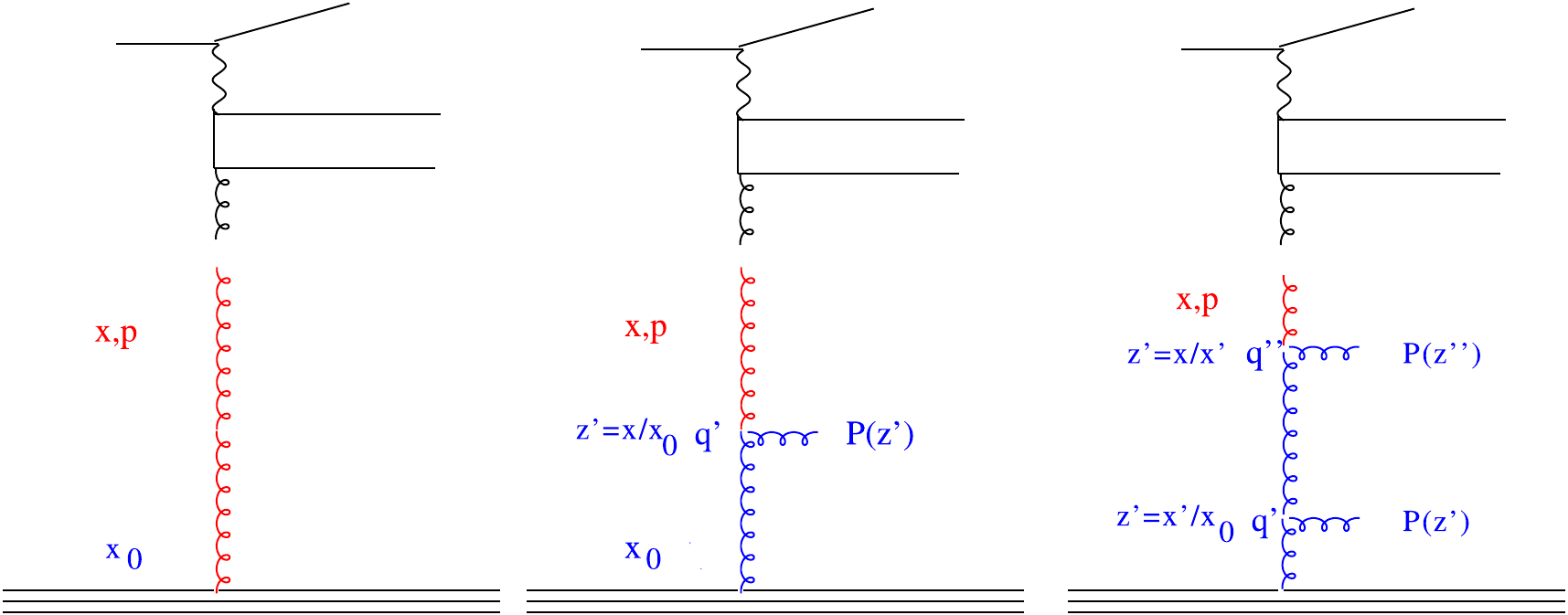}
\caption{Evolution by iteration}
  \label{Fig:evolution}
\end{figure}

In a procedure termed forward evolution, a Monte Carlo  technique~\cite{James:1980yn,Marchesini:1990zy,Marchesini:1992jw,Hautmann:2014uua,Hautmann:2017fcj,Hautmann:2017xtx} is employed to evolve from $\mu_0$ to a value $q'$ determined by the Sudakov factor $\Delta(q')$. This factor represents the probability of evolving from $\mu_0$ to $q'$ without encountering any resolvable branching. 

Let us consider a simplified one-flavor scenario for integrated distribution. Utilizing the Sudakov factor $\Delta$ and the relation
\begin{eqnarray*}
\frac{\partial}{\partial q'}\Delta(q')&=&
-\Delta(q')
\left[\frac{1}{q'}\right]
\int^{\zM} d z {P}^R(z, \as)  , 
\end{eqnarray*}
the  iterative solution of the evolution equation is expressed as  
\begin{eqnarray}
&& xf(x,\mu^2)  =   
 xf ( x , \mu_0^2 )\Delta(\mu^2)  \nonumber \\
 & + &   \int_{\mu^2_0}^{\mu^2} \int_x^1 dz' (-d\Delta(q^{\prime\,2})) \frac{\Delta(\mu^2)}{\Delta(q^{\prime})} {P}^R(z', \as)  { \frac{x}{z'}f(x/z',\mu_0^{\prime 2})) }
  \left[{\int^{\zM} d z {P}^R(z, \as) } \right]^{-1} \nonumber \\ && +...\  .
\end{eqnarray}

The value $q'$ is obtained by solving the equation:
\begin{equation}
R_1 
 =  \int^{q^{\prime\,2}} \frac{\partial \Delta(q^{\prime\prime\,2})}{\partial q^{\prime\prime\,2}}
  dq^{\prime\prime\,2} 
 =  \Delta(q^{\prime\,2}) \ , \label{MC1}
\end{equation}
where $R_1$ is a random number sampled from the interval $[0,1]$.

If $q' > \mu$, the scale $\mu$ is reached, and the evolution halts, leaving only the initial term without any resolvable branching. Conversely, if $q' < \mu$, a branching is generated at $q'$ according to the splitting function $P^R(z', \as)$:
\begin{eqnarray}
\int_{z_{min}}^z  dz' {P}^R(z', \as) & = & R_1 \int_{z_{min}}^{\zM} dz' {P}^R(z', \as) ,  \label{MC2}
\end{eqnarray}
where $R_2$ is a random number in $[0,1]$.

The evolution then continues using the Sudakov factor $\Delta(q'',q')$. If $q'' > \mu$, the evolution stops, leaving a single resolvable branching at $q'$. Otherwise, the evolution continues as described above. This iterative process repeats until $q$'s larger than $\mu^2$ are generated. Through this method, all kinematically allowed contributions in the series are summed, yielding an Monte Carlo estimate of the parton distribution function.

\subsection{Splitting functions, \as , and starting distributions}

The evolution in \updfevolv\ can be carried out at leading order (LO), next-to-leading order (NLO), and next-to-next-to-leading order (NNLO). Splitting functions and \as\ are sourced from \qcdnum~\cite{Botje:2010ay}, while the initial parton density distributions are obtained in the same format. The NLO and NNLO splitting functions are calculated in Refs~\cite{\nnloSplit}.

Alternatively, LHApdf~\cite{Buckley:2014ana} offers an alternative source for the starting distributions, as well as for the parametrization of \as .

 \subsection{Computational Techniques: \updfevolv\ Grid}

In fitting programs where the DGLAP evolution is employed to determine the starting distribution ${\cal A}_0 (x)$, a full Monte Carlo solution~\cite{Hautmann:2014uua} becomes impractical due to its time-consuming nature and susceptibility to numerical fluctuations. Instead, a convolution method introduced in~\cite{Jung:2012hy,Hautmann:2013tba} is utilized.

The kernel $ {\cal K}\left(x'',\kt,\Pmax\right) $, where $|{\bf k} | = \kt $,  is determined from the Monte Carlo solution of the PB evolution equation and then convoluted with the non-perturbative starting distribution ${\cal A}_0 (x)$:

\begin{eqnarray}
x {\cal A}(x,\kt,\Pmax) &= &x\int dx' \int dx'' {\cal A}_0 (x') {\cal K}\left(x'',\kt,\Pmax\right)
 \delta(x'
x'' - x) \nonumber \\
& = & \int dx' {{\cal A}_0 (x') } 
\cdot \frac{x}{x'} \ { {\cal K}\left(\frac{x}{x'},\kt,\Pmax\right) } .
\end{eqnarray}

The kernel ${\cal K}$ encapsulates all dynamics of the evolution, including Sudakov form factors and splitting functions. It is determined on a grid of $50\otimes50\otimes50$ bins in $ x,  \kt,  \Pmax$. The grid's binning is logarithmic, with 40 bins in logarithmic spacing below 0.1 for the longitudinal variable $x$, and 10 bins in linear spacing above 0.1.

The starting distributions can be obtained from fits to measurements via the xFitter platform~\cite{xFitterDevelopersTeam:2022koz,Alekhin:2014irh}. The user has to provide the grid files for gluons, as well as for u- and d-type quarks obtained from \updfevolv, both collinear and TMD grids are needed. These grid files are being used inside xFitter convoluted with the starting distributions to provide evolved parton distribution functions (pdfs).

After a fit is performed, the resulting collinear pdfs are written in LHApdf format, the TMD parton densities are written in TMDlib format~\cite{Abdulov:2021ivr}. Both collinear and TMD parton distributions can be plotted using the graphical web interface \TMDplotter ~\cite{Abdulov:2021ivr}.

\subsection{Determination of electroweak particle densities}
The DGLAP evolution equation can be extended to include also photons and other  electroweak particles \cite{Fornal:2018znf,Bauer:2017isx,Veness-2012,Ciafaloni:2005fm,Ciafaloni:2001mu}. Since the quarks carry different electric and weak charges, it is necessary to split the evolution into u-type and d-type quarks.

\begin{figure}[h!tb]
\begin{center} 
\vskip -3cm
\includegraphics[width=0.8\textwidth,angle=270]{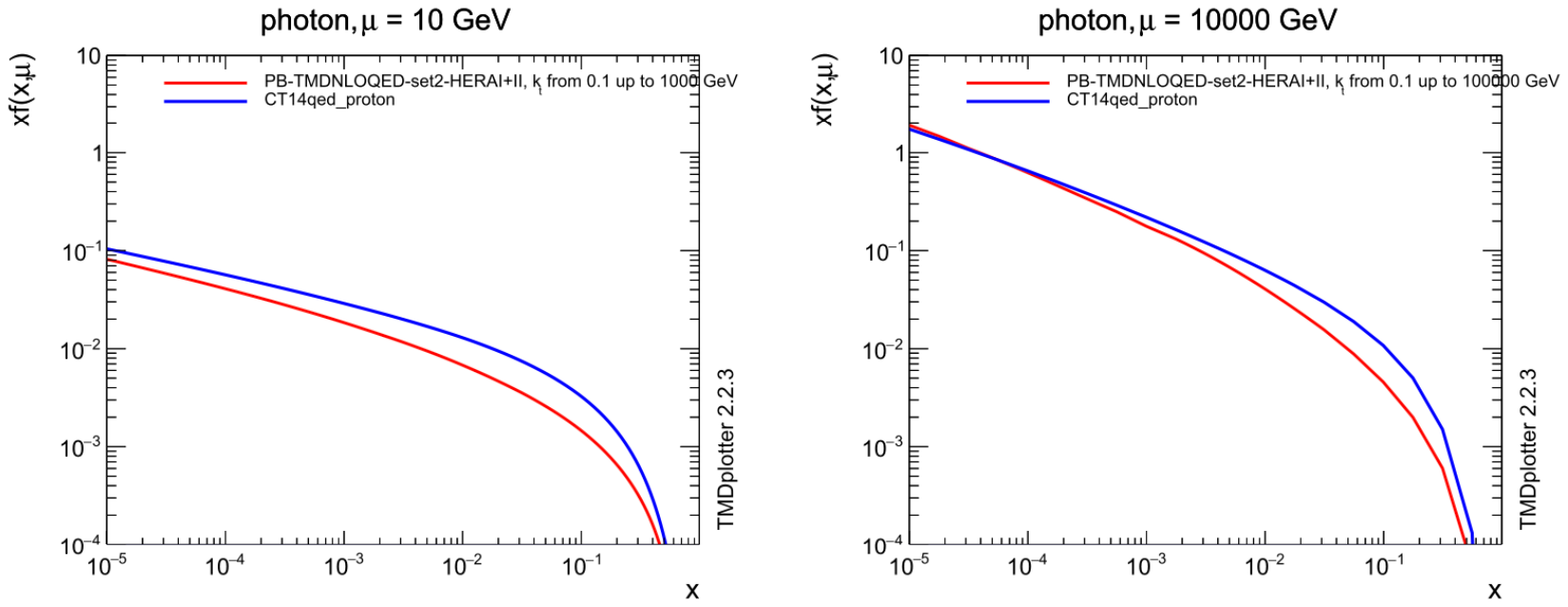}
\vskip -3cm
\caption{\small The collinear photon density  at $\mu= 10$ GeV and $\mu= 10000$ GeV as a function of $x$. For comparison  the prediction from  CT14qed-proton~\protect\cite{Schmidt:2015zda} is also shown.}
\label{PhotonFig1}
\end{center}
\end{figure}

The evolution of the photon density inside a hadron has been described in Ref.~\cite{Jung:2021mox,Wening:449805}.
In Figure~\ref{PhotonFig1} the collinear photon density as a function of $x$ is shown for different values of the evolution scales $\mu$ (from Ref.~\cite{Jung:2021mox}). For comparison also the prediction from CT14qed-proton~\cite{Schmidt:2015zda} is shown.

\begin{figure}[h!tb]
\begin{center} 
\vskip -2.5cm
\includegraphics[width=0.7\textwidth,angle=270]{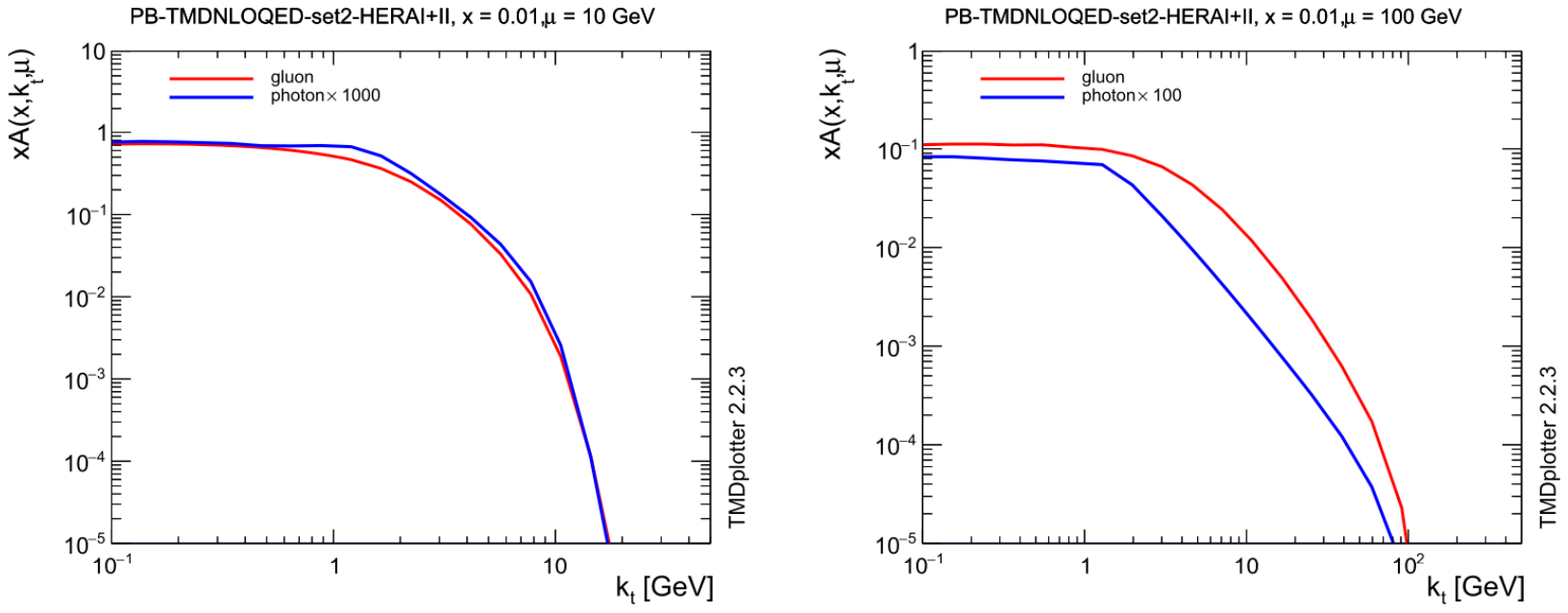}
\vskip -3cm
\caption{\small The TMD photon density  at $\mu= 10$ GeV and $\mu= 100$ GeV as a function of $\kt$. For comparison the gluon density is also shown. Plot from \protect\cite{Jung:2021mox}.}

\label{PhotonFig12}
\end{center}
\end{figure}

In Figure~\ref{PhotonFig12} the TMD density of photons is shown (from Ref.~\cite{Jung:2021mox}).

The determination of effective $\rm{W}$ densities has been discussed already in Refs.~\cite{Kane:1984bb,Lindfors:1985yp,Cahn:1984tx,Dawson:1984gx,Chanowitz:1985hj,Kleiss:1986xp,Dawson:1986tc,Altarelli:1987ue,Kunszt:1987tk}. In recent years, this ideas has been picked up again in Refs.~\cite{Ciafaloni:2005fm,Bauer:2017isx,Bauer:2017bnh,Fornal:2018znf,Ciafaloni:2024alq}.

\begin{figure}[h!tb]
\begin{center} 
\includegraphics[width=0.45\textwidth,angle=0]{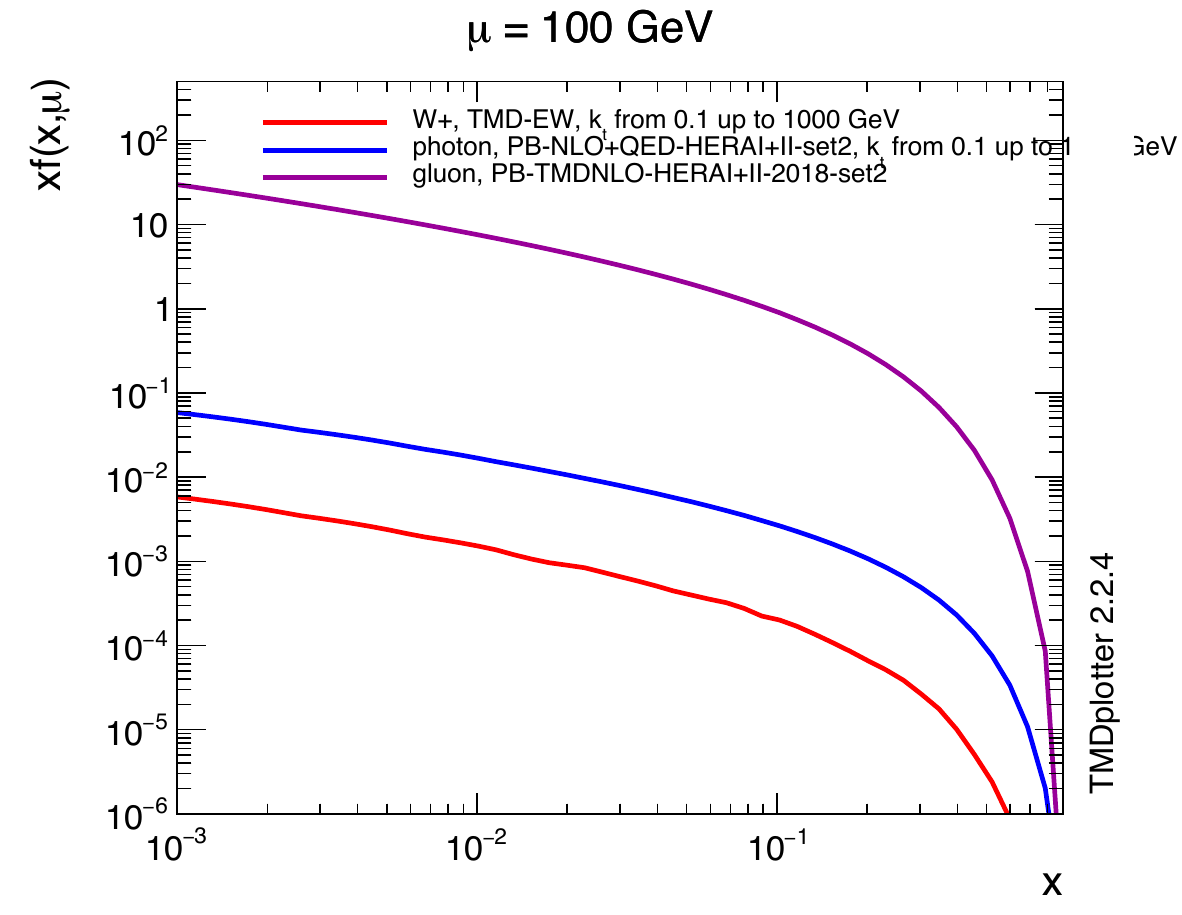}
\includegraphics[width=0.45\textwidth,angle=0]{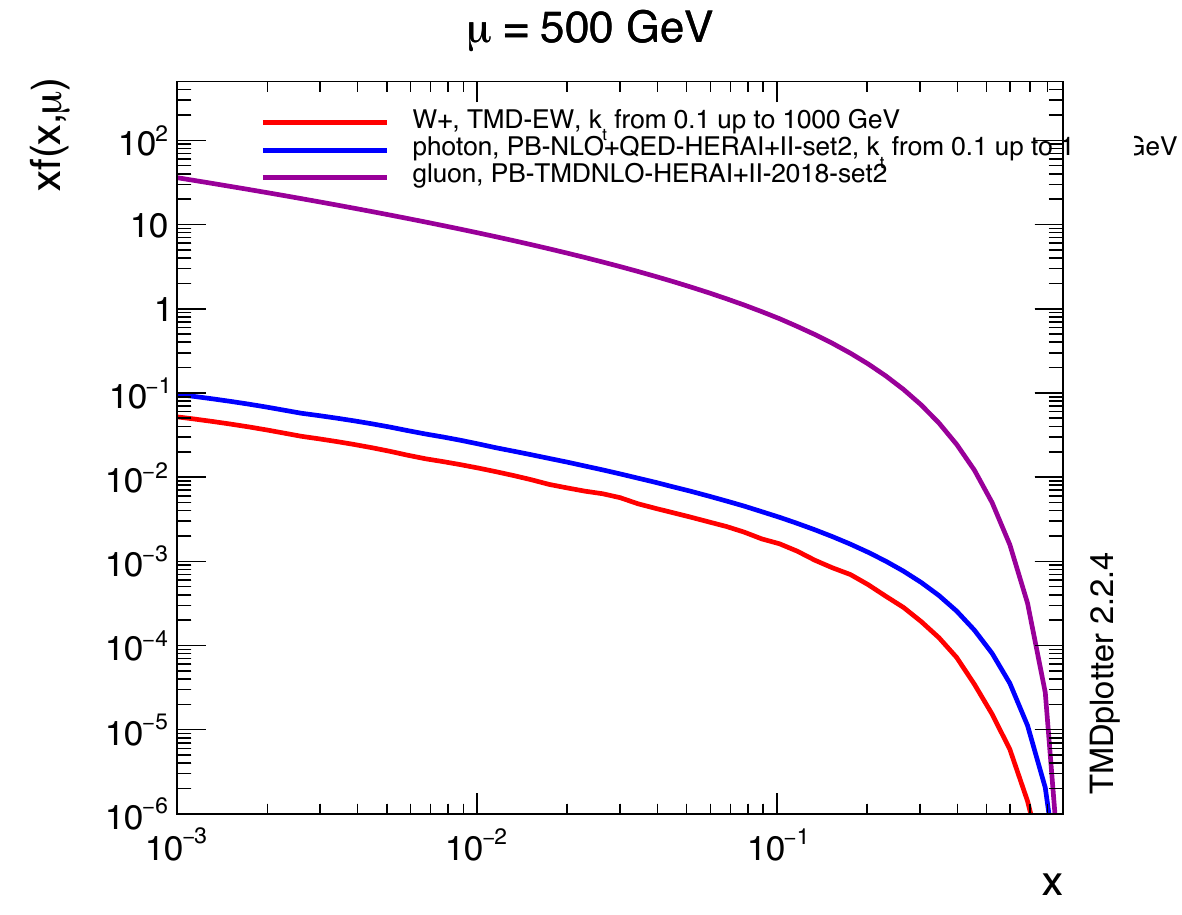}
\caption{\small The collinear vector boson densities  at $\mu= 100$ GeV and $\mu= 500$ GeV as a function of $x$.}
\label{EWFig1}
\end{center}
\end{figure}

\begin{figure}[h!tb]
\begin{center} 
\includegraphics[width=0.45\textwidth,angle=0]{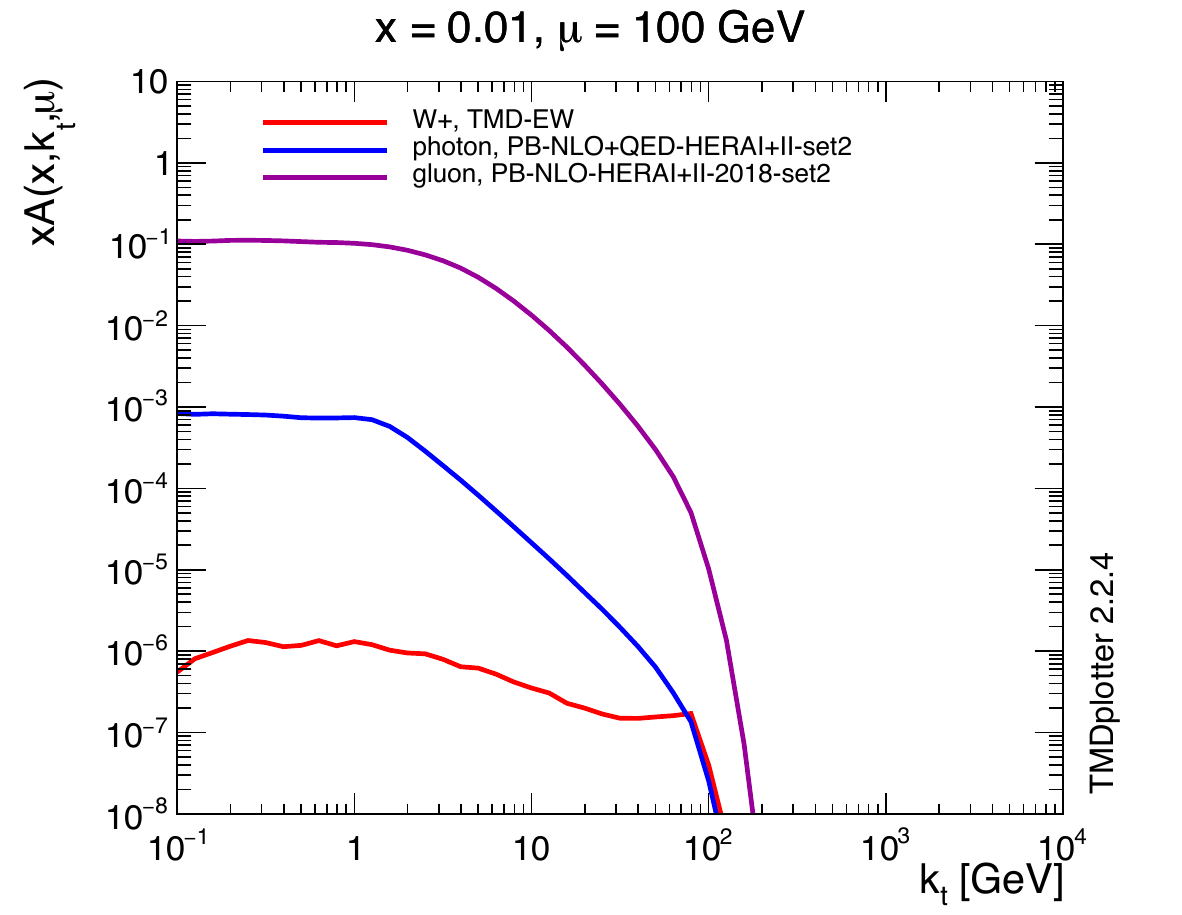}
\includegraphics[width=0.45\textwidth,angle=0]{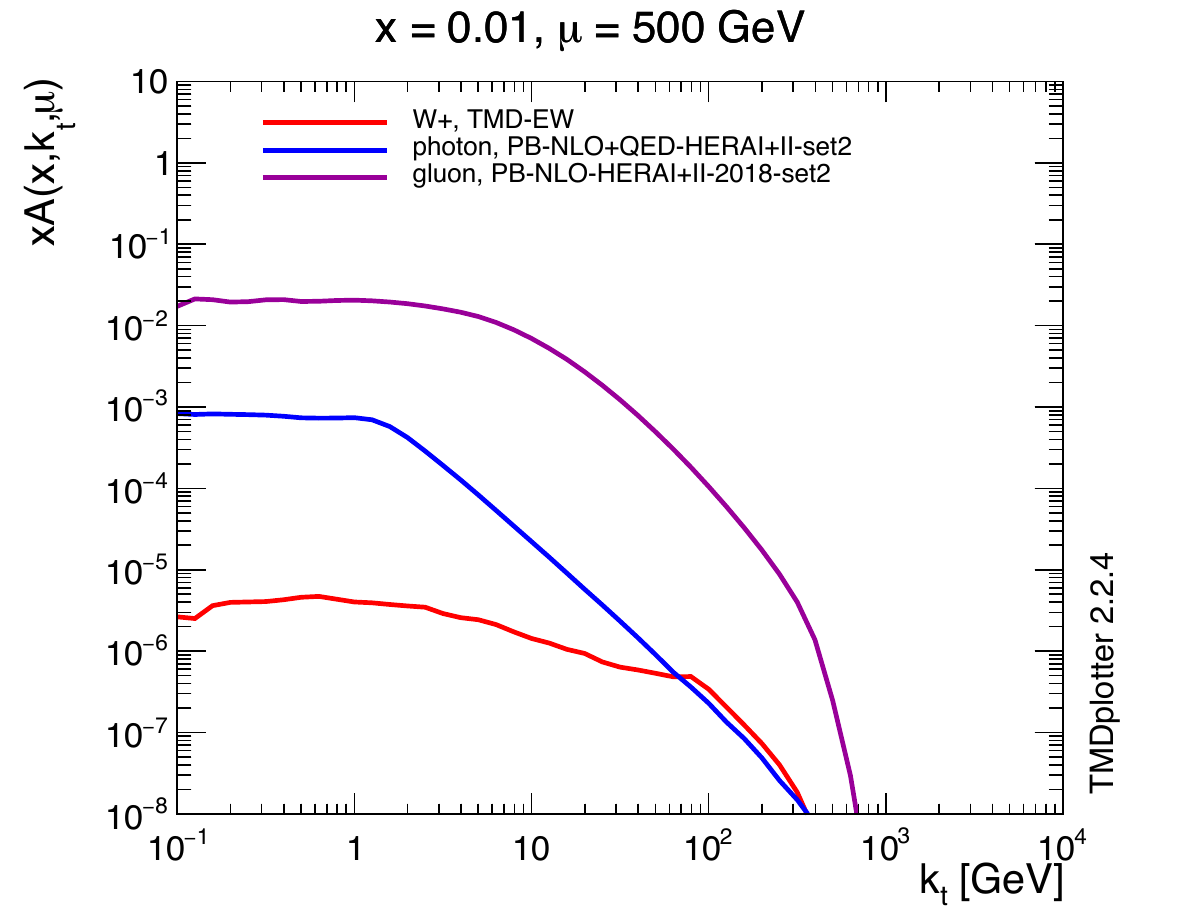}
\caption{\small The TMD vector boson densities  at $\mu= 100$ GeV and $\mu= 500$ GeV as a function of \kt .}
\label{EWFig2}
\end{center}
\end{figure}

The approach to determine the photon densities within the \PBM -method can be easily extended to calculate the collinear and TMD densities of $\rm{Z}$ and $\rm{W}$. The straightforward application of the method gives the collinear and TMD densities as shown in Figures~\ref{EWFig1},\ref{EWFig2}, compared also to the photon density.

The heavy vector-boson density vanishes (in the approach applied here) for scales $\mu < m_{\rm{W}} $ therefore the densities are only shown for larger scales. For higher scales, the photon and $\rm{W}$ densities approach each other, as they should.

In the transverse momentum distribution (Figure \ref{EWFig2}), one can observe the similarity of photon and $W$ densities at large \kt , while significant differences are visible at smaller \kt .

\subsection{Application: Predictions using \PBM\ collinear and TMD distributions}
The  PB TMDs were fitted to the DIS precision data~\cite{Abramowicz:2015mha}, as described in detail in Ref. \cite{Martinez:2018jxt}.

The CASCADE3 Monte Carlo generator \cite{Baranov:2021uol}, engineered specifically to conform to the PB TMD approach, is the only generator with parton shower based on TMDs. Special strength of this approach is that CASCADE3 allows to simulate parton showers fully consistent with TMD parton distribution functions. The PB method and CASCADE3 provide successful predictions for Deep-Inelastic Scattering (DIS) \cite{Martinez:2018jxt,H1:2021wkz}, inclusive Drell-Yan (DY) at different $\sqrt{s}$ and mass ranges \cite{BermudezMartinez:2020tys,CMS:2022ubq,Martinez:2019mwt,Bubanja:2023nrd}, Z+jets \cite{Yang:2022qgk,CMS:2022ilp} and dijets \cite{CMS:2022drg,Abdulhamid:2021xtt}. Furthermore, the PB method has also been included in various publications of CMS \cite{CMS:2022vkb,CMS:2022drg,CMS:2021lxi} and H1 \cite{H1:2021wkz} showing a great potential for becoming commonly used in the experimental community. 


\section{Description of the program components}
%
%
\subsection{Subroutines and functions}

The source code of \updfevolv\ and this manual can be found under:\\
\verb+https://updfevolv.hepforge.org/+

\begin{defl}{123456789012345678}
\item[{\tt      sminit}] to initialise
\item[{\tt      sminfn}] to generate starting distributions in $x$ and $k_T$
\item[{\tt      smbran}] to simulate perturbative branchings
\item[{\tt      szval}] to calculate $z$ values for the splitting
\item[{\tt      smqtem}] to generate $q^{\prime}$ from the corresponding Sudakov factor
\item[{\tt      updfgrid}] to build, fill and normalise the updf grid.
\item[{\tt      asbmy(q)}]    to calculate $\frac{ \as (q)} { 2 \pi}$ at LO, NLO or NNLO using the utility from \qcdnum\ or to take it from LHApdf.
\\

\item[{Utility routines:}]
\item[{\tt evolve tmd}] Main routine to perform parton evolution
\item[{\tt updfread}] example program to read and plot the results
\item[{\tt  gadap}] 1-dimensional Gauss integration routine	
\item[{\tt  gadap2}] 2-dimensional Gauss integration routine
\item[{\tt  divdif}]      linear interpolation routine  (CERNLIB)
\item[{\tt      ranlux}]  Random number generator \verb+RANLUX+ (CERNLIB)
\end{defl}

\subsection{Parameter in steering files}
\begin{defl}{123456789012345678}
\item[]
\item[{\tt  Ipdf = 60500}] set name for collinear valence quark starting distribution
\item IPDF=-1: QCDNUM parametrization  used as the starting distribution
\item IPDF$>$0: LHApdf set name for the starting distribution
\item[{\tt  Qg = 1.3}] starting value $\mu_0$ for perturbative evolution
\item[{\tt  Qs = 1.0}]  Gaussian width for the intrinsic $k_T$ distribution 
\item[{\tt  Iordas = 2}]   order in perturbation theory for splitting functions   (iordas = 1 is LO, iordas=2 is NLO,  iordas=3 is NNLO). If QCDNUM is used as a starting distribution, Iordas defines also the order of $\alpha_s$ (for IPDF>0,  $\alpha_s$ is accessed from LHApdf)
\item[{\tt  Iqord = 2}]   ordering definition to calculate scale in $\alpha_s$, the transverse momentum \qt\ and $z_M$.
\item Iqord=0 : Angular ordering with the scale of $(1-z)^2 q^{\prime\,2}$  in $\alpha_s$ and  $z_M=z_{\rm{dyn}} = 1-q_0/q'$
\item Iqord=1 : Virtuality ordering with the scale of $(1-z)q^{\prime\,2}$  in $\alpha_s$ and  $z_M=z_{\rm{dyn}} = 1-(q_0/q')^2$
\item Iqord=2: Angular ordering with the scale of $q^{\prime\,2}$  in $\alpha_s$ and fixed \zM
\item[{\tt  zmaxfixed = 0.999999}] active only for Iqord=2
\item[{\tt  Nev = 1000000}] Number of generated events
\item[{\tt  Q0ord=0.01}] active only for Iqord=0 and 1\\
$q_0$ value for dynamical $z_{M}=z_{\rm{dyn}}$ 
\item[{\tt  Ikern=0}]  
\item Ikern=0 : full evolution with starting distribution (run ends up with two grid files for TMD and collinear distribution)
\item Ikern=1 : only kernel to be used in xFitter (run ends up with four grid files, two for gluon and quark TMD and two collinear kernels)
\item[{\tt  mc=1.47}] charm mass 
\item[{\tt  mb=4.5 }] bottom mass   
\item[{\tt  mt=173.}] top mass 
\item[{\tt  asZ=0.118}] $\as(m_Z)$   
\end{defl}

\subsection{Storing of the outputs}
\begin{defl}{123456789012345678}
\item[{\tt  updf-grid.dat}] name of the grid file for TMD outcome.
\item[{\tt  updf-grid\_int.dat}] name of the grid file for collinear outcome.
\item[{\tt  test.root}] name of the root file containing histograms with collinear pdfs.
\end{defl}

In order to have enough statistics, usually 900 jobs with $10^6$ events are needed. The results of each job are then added to produce the final grid files. The code for doing this can be obtained from the authors.
\section{Program Installation}
 \updfevolv\ follows the standard AUTOMAKE
convention. To install the program, do the following
\begin{tiny}
\begin{verbatim}
1) Get the source

tar xvfz uPDFevolv2-XXXX.tar.gz
cd uPDFevolv2-XXXX

2) Generate the Makefiles (do not use shared libraries)
./configure

3) Compile the binary
make

4) Install the executable
make install

4) The executable is in bin

run it with:
bin/updf_evolve < steering.txt

plot the result with:

bin/updfread

\end{verbatim}
\end{tiny}

 \section{Acknowledgments}
 We are grateful to R. Zlebcik and L. Keersmaekers for many discussions during the evolution of the \updfevolv\ code.
 A. Lelek acknowledges funding by Research Foundation-Flanders (FWO) (application number: 1272421N).

\bibliographystyle{mybibstyle-new.bst}
\raggedright  
\providecommand{\href}[2]{#2}\begingroup\raggedright\endgroup

\end{document}